\documentclass[doublecol]{epl2} 
\usepackage{color}
\usepackage[english]{babel}
\usepackage[latin1]{inputenc}
\usepackage{graphicx,times}
\usepackage{amsmath,amssymb}
\usepackage{subfigure}
\usepackage{array}

\providecommand{\e}[1]{\ensuremath{\cdot 10^{#1}}}

\title{The structure of coevolving infection networks}

\author{Stefan Wieland\inst{1}  \and Tom\'as Aquino\inst{1,2} \and Ana Nunes\inst{1} \thanks{E-mail: \email{anunes@ptmat.fc.ul.pt}}}
\shortauthor{Wieland \etal}

\institute{                    
  \inst{1} Centro de F{\'\i}sica da Mat\'eria Condensada and
Departamento de F{\'\i}sica, Faculdade de Ci{\^e}ncias da Universidade de 
Lisboa, P-1649-003 Lisboa, Portugal\\
  \inst{2} Centro de Matem\'atica e Aplica\c c\~oes Fundamentais \\
Faculdade de Ci{\^e}ncias da Universidade de Lisboa, P-1649-003 Lisboa, Portugal
}
\pacs{05.10.Gg}{Stochastic analysis methods}
\pacs{87.10.Mn}{Stochastic modeling}
\pacs{89.75.Fb}{Structures and organization in complex systems}

\abstract{Disease awareness in infection dynamics can be modeled with adaptive contact networks whose rewiring rules  reflect the attempt by susceptibles to avoid infectious contacts. Simulations of this type of models show an active phase 
with constant infected node density in which 
the interplay of disease dynamics and link rewiring prompts the convergence towards a well defined degree distribution, irrespective of the initial network topology. We develop a method to study this dynamic equilibrium and give an analytic description of the structure of the characteristic degree distributions and other network measures. The method applies to a broad class of systems and can be used to determine the steady-state topology of many other adaptive networks.}

\begin{document}

\maketitle

\section{Introduction}

Research on the influence of network topology on the global dynamics of systems composed of many interacting units has been very active for the last ten years. More recently, adaptive and coevolving networks, that is, networks whose topology changes along with the dynamics, have begun to be considered  on two standings. The first one is to explore the consequences of this interplay from the viewpoint of global dynamics \cite{general_refs2}. The second is to understand what sort of interactions may give rise to certain network architectures  \cite{general_refs1}. Understanding the origin of degree distributions and correlations of real networks was first addressed by considering networks  that evolve independently of the dynamics \cite{Barabasi-Albert}, and it is natural to extend this idea to networks whose evolution is determined by the dynamics they support.  \par \noindent

Infection dynamics is a natural setting to investigate the coupled evolution 
of the 
interacting elements and the network of interactions, because disease awareness translates into 
susceptibles that try to evade infection by changing their contact patterns 
according to the disease status of their neighbors.  The SIS (susceptible, infected, susceptible) and the SIR (susceptible, infected, recovered) models are the simplest models for the spread of epidemic infections \cite{Murray}. They are based on the idea that an infected individual infects susceptibles at a certain rate for a certain period, and then recovers and becomes susceptible again (if the disease does not confer immunity), or recovers and becomes immune. Stochastic versions of these models on networks can be mapped onto bond percolation \cite{bondperc}. \par \noindent

\begin{figure}
\centering
  \includegraphics[width=8.6cm]{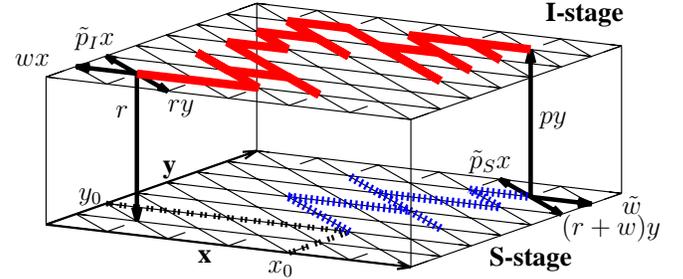}
  \caption{(Color online) Status and joint degree evolution of a node going through the S-stage (blue dashed line) and I-stage (red solid line), starting with joint degree \((x_0,y_0)\) in the S-stage. Bold arrows show all possible transitions and their rates.}
\label{f:RW}
\end{figure}

Simple epidemic models like these exhibiting trivial global dynamics on static networks have been shown to display interesting behavior on adaptive networks, in which links between nodes may be created or removed according to rewiring rules dictated by disease awareness.
In the network scheme of SIS dynamics, susceptible nodes ({\emph{S-nodes}, a fraction 
\([S]\) of the total number of nodes) are infected with rate \(p\) along links connecting them to infected nodes ({\emph{I-nodes}, the remaining fraction \([I]=1-[S]\)), whereas the latter recover with rate \(r\). Gross et al. \cite{Gross} incorporated disease awareness into the conventional SIS model by introducing a topology-changing, yet link-number preserving, rewiring mechanism:
S-nodes try to evade infection by retracting links from infected neighbors with rate \(w\), and rewiring them to randomly selected S-nodes. The description of the model
in the framework of the standard pair approximation yields a complex phase diagram. 
In addition to a frozen phase (where the disease dies out) and to a stationary active phase (where \([I]\neq 0\) is constant and \([AB]\), the number per node of links
connecting nodes of type \(A\) and \(B\), \(A,B \in \{S,I\}\), is also constant), 
other active phases emerge that are absent in the SIS model without rewiring, namely an oscillatory phase and a bistable phase \cite{Gross2}.} 
Simulations in the stationary active phase show the states of the nodes and the links coevolving  to produce and maintain a dynamic network topology characterized by well defined degree distributions not only for the global network, but also for the subsets of 
S- and I-nodes. This remarkable fact and the structure of the characteristic degree distribution were left mostly unexplained.
In the wake of this first paper based on SIS infection dynamics, several contributions have addressed extensions and modifications \cite{adaptiveSIX} of the original model, focusing mostly 
on the global dynamics and the conditions for disease persistence. In a recent paper \cite{quebec},
steady-state degree distributions from individual-based simulations were shown to be very well approximated by the outcome of long-term numerical integration of a compartmental
model for the evolution of the node status together with the number of its infected and susceptible neighbors. This model provides an alternative to simulations to produce approximately the observed network topology, but it is not amenable to analytic treatment. 
Overall, the description of the network topology underlying the steady states of these models has received comparatively little attention. Here we develop
an analytic method to compute degree distributions and other measures that characterize this network topology.  
\par \noindent

\section{The node-cycle model and its solution} 

We illustrate the method taking the original model \cite{Gross} as an example of an adaptive network, and focusing on the stochastic process that each node and its links follow as the former undergoes status change from susceptible to infected, and gains (in the susceptible \emph{S-stage}) or loses (in the infected \emph{I-stage}) links. This process can be treated analytically, and the stationary degree distributions for the full \emph{node cycle} (NC), from susceptible to infected and back again, can be obtained and 
compared with the degree distributions of the susceptible and infective subnetworks of the original network model. The NC approach is based on the ergodic properties of the network stationary state, which are a consequence of the random rewiring process that ensures mixing. Indeed, we observed in simulations of the steady state of the network model that recording ensemble statistics yields the same distributions as sampling a single node over a sufficiently long time. 
\par \noindent

In each stage of the NC we keep track of the numbers 
\(x\) and \(y\) of susceptible and infected neighbors of the node (its \emph{joint degree} (x,y)).
Due to infection and rewiring processes, these numbers change at
certain rates, defining a time-homogeneous Markov process that can be described in each stage as a finite random walk on a degree grid spanned by \(x\) and \(y\) (Fig.~\ref{f:RW}).
The time-continuous random walks performed in 
each stage are one-step processes in both coordinates. They are coupled by stage transition, which takes place at constant rate
\(r\) from the I-stage to the S-stage, and at rate \(p\, y\) from the S-stage to the I-stage.
The probability \([x,y]\equiv P_S(x,y,t|x_0,y_0)\) of a walker in the S-stage, having started at coordinates \((x_0,y_0)\), being at \((x,y)\) at time \(t\) obeys the Master Equation 
\begin{align}\label{e:me}
\frac{d[x,y]}{d t}  =& (w+r)\{\left(y+1\right)[x-1,y+1]-y[x,y]\} \nonumber\\ 
& - p\ y[x,y] +\tilde{w}\left([x-1,y]-[x,y]\right)\nonumber \\  
& +\tilde{p}_S\{\left(x+1\right)[x+1,y-1]-x[x,y]\}
\end{align}
with the boundary conditions \([-1,y]=[x,-1]=0\).
The first term on the right hand side of Eq.~(\ref{e:me}) takes into account the swapping of an infected neighbor with a susceptible neighbor due either to rewiring or to recovery of the infected neighbor. The second term on the right represents the transition
to the I-stage and an overall probability mass loss as in the active phase all susceptible nodes
eventually become infected}.
The stochastic dynamics of the local random variables \(x\) and \(y\) is coupled to the global network dynamics through the total degree gain rate \(\tilde{w}\) and through the force of infection \(\tilde{p}_S\) that determines the infection rate of susceptible neighbors of the node.
These two transitions are represented by the third and fourth terms on the right-hand side of Eq.~(\ref{e:me}). A similar Master Equation holds for the I-stage, where instead of \(\tilde{p}_S\) the rate \(\tilde{p}_I\) accounts for the force of infection on susceptible neighbors of an I-node. We assume the \emph{correspondence parameters} \(\tilde{w}\),  \(\tilde{p}_S\) and \(\tilde{p}_I\) to be constant and will assign values later on.

The Master Equation (\ref{e:me}) can be solved using the generating function formalism \cite{vanKampen}.
Multiplying both sides of Eq.(\ref{e:me}) by \( \chi^x \gamma^y\) and summing over \(x\) and \(y\) yields for the probability generating function
\begin{equation*}
F_S(\chi,\gamma,t|x_0,y_0)=\sum\limits^{\infty}_{x,y=0}\chi^{x}\gamma^{y} P_S(x,y,t|x_0,y_0)
\end{equation*}
a linear first order PDE 
\begin{align}\label{e:pde}
\frac{\partial F_S}{\partial t} = & \{ (w+r)\chi-(w+r+p)\gamma \} \frac{\partial F_S}{\partial \gamma}  \nonumber\\
& +\tilde{p}_S (\gamma-\chi) \frac{\partial F_S}{\partial \chi} + \tilde{w}(\chi-1)F_S \, .
\end{align}
Its solution, given that \(F_S(\chi,\gamma,0|x_0,y_0)=\chi^{x_0}\gamma^{y_0}\), is
\begin{align}\label{e:Fs}
 F_S(\chi,\gamma,t|x_0,y_0) =& e^{-\tilde{w}\{c_5\left(t\right)\chi+c_6\left(t\right)\gamma+c_7t\}}\nonumber \\
& \times \{c_1\left(t\right)\chi+c_2\left(t\right) \gamma\}^{x_0}\nonumber \\
& \times \{c_3\left(t\right)\chi+c_4\left(t\right)\gamma\}^{y_0} \, ,
\end{align}
where \(c_j(t), j\in \{1,...,6\}\) consist of linear combinations of exponential functions of time and \(c_7>0\).

The Taylor expansion of Eq.~(\ref{e:Fs}) up to arbitrary order is easy to compute, yielding \( P_S(x,y,t|x_0,y_0)\). With the remaining probability mass at time \(t\) being \(F_S(1,1,t|x_0,y_0)\), the total occupation time at \((x,y)\), having started at \((x_0,y_0)\), reads as
\begin{equation*}\label{e:jdd0}
P_{S}(x,y|x_0,y_0)= 
 \int \limits_{0}^{\infty}P_{S}(x,y,t|x_0,y_0)\, {\upd t} \, .
\end{equation*}
Given that the rate of reinfection of a S-node with \(y\) infected neighbours is \(p \ y\),
we calculate
\begin{equation*}
\hat{P}_S(x,y|x_0,y_0)= p \, y \int \limits_{0}^{\infty} P_S(x,y,t|x_0,y_0) \, {\upd t} 
\end{equation*}
to obtain the probability of, having started at \((x_0,y_0)\), ending the walk in the S-stage at \((x,y)\).
In a similar way, a closed-form expression for the I-stage can be obtained, where \(\hat{P}_I=r P_I\) because of recovery being coordinate-independent.

The elements of \(\hat{P}_A\), \(A \in \{S,I\}\), are the transition probabilities that map 
a random walker's initial coordinate distribution (ICD) in stage \(A\), 
\(\Phi_A(x,y)\), onto the ICD of the subsequent stage according to
\begin{equation}\label{e:switch}
\Phi_{I,S}(x,y)=\sum\limits_{x_0,y_0=0}^{\infty} \hat{P}_{S,I}(x,y|x_0,y_0)\,\Phi_{S,I}(x_0,y_0).
\end{equation}
The ICD of the S-stage after a full node cycle changes according to
\begin{equation*}
\Phi_{S}'(x,y) =\sum\limits_{x_0,y_0=0}^{\infty}\hat{P}{}(x,y|x_0,y_0)\,\Phi_S(x_0,y_0),
\end{equation*}
where the \(\hat{P}(x,y|x_0,y_0)\) are given by the Chapman-Kolmogorov identity
\begin{equation*}
\hat{P}(x,y|x_0,y_0) \equiv\sum\limits_{x',y'=0}^{\infty} \hat{P}_I(x,y|x',y') \hat{P}_S(x',y'|x_0,y_0).
\end{equation*}
Setting a cutoff for the maximum degree, the elements of \(\hat{P}\) encode a finite ergodic Markov chain, and therefore an arbitrary ICD in the S-stage will converge upon iteration of the node cycle to a unique stationary state \(\Phi_{S}^*(x,y)\) which, according to the Perron-Frobenius theorem, can be found as the right eigenvector of \(\hat{P}\) associated with the eigenvalue \(1\). Once \(\Phi_{S,I}^*(x,y)\) are known, the normalized total occupation times of coordinates in the stationary state can be computed in either stage as
\begin{equation}\label{e:jdd}
P^*_{S,I}(x,y)=\frac{\sum\limits_{x_0,y_0=0}^{\infty} P_{S,I}(x,y|x_0,y_0)\,\Phi_{S,I}^*(x_0,y_0)}
	{\sum\limits_{x_0,y_0=0}^{\infty}\ \int\limits_{t=0}^{\infty} F_{S,I}(1,1,t|x_0,y_0)\mathrm{d}t \ \Phi_{S,I}^*(x_0,y_0)}\, ,
\end{equation}
where in particular
\begin{equation*}
P_I^*(x,y)=\Phi_S^*(x,y)
\end{equation*}
because of Eq.~(\ref{e:switch}) and \(\hat{P}_I=r P_I\). 
In each stage $A$, $A \in \{S,I\}$, the average number of infected and susceptible neighbors, $\langle x_A\rangle $ and $\langle y_A\rangle $, and the average degree, $\langle k_A\rangle = \langle x_A\rangle  + \langle y_A\rangle $,  can be computed from the distributions $P_{S,I}^*(x,y)$.

The lifetime distributions in each stage \(T_{S,I}(t,x,y)\) of random walkers with initial coordinates \((x,y)\) and the generating functions \(F_{S,I}\) are related through
\begin{equation*}
F_{S,I}(1,1,t|x,y)=1-\int \limits_{0}^{t} T_{S,I}(t',x,y) \,\upd t' \, ,
\end{equation*}
with overall lifetime distributions $T_{S,I}(t)$
\begin{equation*}
T_{S,I}(t)= \sum\limits_{x,y=0}^{\infty}\Phi^*_{S,I}(x,y) T_{S,I}(t,x,y)
\end{equation*}
and survival functions \(L_{S,I}(t)\)  in the steady state
\begin{align}\label{e:LT}
L_{S,I}(t)&=1-\int \limits_{0}^{t}T_{S,I}(t')\, \upd t' \nonumber \\
&=\sum\limits_{x,y=0}^{\infty}\Phi^*_{S,I}(x,y) F_{S,I}(1,1,t|x,y) \, .
\end{align}
The average duration \(\tau_{S,I}\) of each stage in the stationary state is thus given by 
\begin{equation*}
\tau_{S,I} \ = \sum\limits_{x,y=0}^{\infty} \Phi_{S,I}^*(x,y) \int \limits_{0}^{\infty } F_{S,I}(1,1,t|x,y) \,\upd t \,,
\end{equation*}
It then follows from the solution of the analogue of Eq.~(\ref{e:pde}) for the I-stage that \(\tau_{I} =1/r\) as expected, because there random walkers switch to the S-stage at a coordinate-independent constant rate \(r\).

\section{Correspondence with the network model and comparison with simulations}

The stationary state of the NC process was solved analytically for any choice of the parameters \(w\), \(r\), \(p\) and \(\tilde{w}\),  \(\tilde{p}_S\), \(\tilde{p}_I\).
These solutions encompass a much larger set of systems than the original network model of \cite{Gross}, because the correspondence parameters, tying network to NC dynamics, have  yet to be determined. 
A first estimate for these parameters may be obtained from the equilibrium link number per node predicted by the pair approximation model of \cite{Gross}. Combining this approximation with the NC description yields $\tilde{w} = w\ [SI]/(1-[I])$,  and similar expressions hold for the other correspondence parameters. The results for the subensemble degree distributions obtained through this approach reproduce the qualitative behaviour of the network simulations but the quantitative agreement is poor (results not shown).
An alternative way of constraining the correspondence parameters comes out of embedding 
the NC description in a given network of fixed degree. On one hand, the correspondence parameters must be such that link depletion in the I-stage and link addition in the S-stage balance out for that degree. On the other hand, they should ensure that computing the  numbers of links between nodes of type \(S\) and \(I\) does not depend on whether one sums over the infected neighbors of all the S-nodes or over the susceptible neighbors of all I-nodes. These consistency conditions will provide a self-contained
determination of the correspondence parameters within the NC framework which is independent of the pair approximation and does not involve network simulations for parameter fitting.

For the NC process to represent the stationary dynamics of the network, so that the \(P_{S,I}^*(x,y)\) correspond to the network subensemble degree distributions and \( \tau_S\) and \( \tau_I\) are proportional to the 
fractions \([I]\) and \([S]\) of infected and susceptible  nodes in the network,  
we shall therefore impose the following two conditions:

\noindent
(i) The average node degree must be equal to the average degree $\langle k \rangle$ that is chosen a priori in the network model, 
\begin{equation*}
\tau_S \langle k_S\rangle + \tau_I \langle k_I\rangle = \langle k \rangle (\tau_S + \tau_I ), 
\end{equation*}
where  \(\langle k_{S,I}\rangle \) are 
the average degrees computed from Eq.~(\ref{e:jdd}).

\noindent
(ii) The S- and I-stages represent nodes that are each other's neighbors, and therefore 
 \(\tau_S \langle y_S\rangle = \tau_I \langle x_I\rangle \) for the mean number of susceptible and infected neighbors in the two stages, with the averages again computed from Eq.~(\ref{e:jdd}).

\begin{figure}
\centering
  \includegraphics[width=8.6cm]{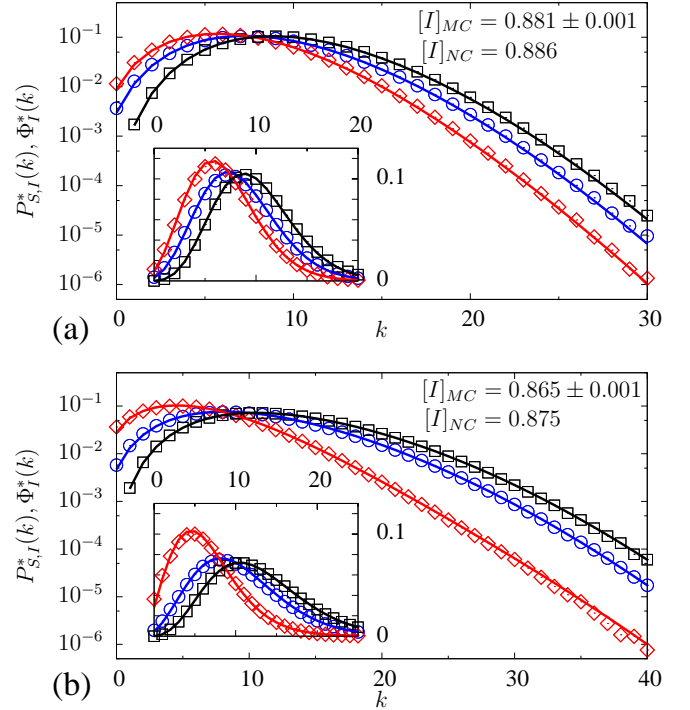}
  \caption{
(Color online) Subensemble steady-state degree distributions for susceptibles (circles), infected (diamonds) as well as the steady-state ICD for the I-stage (squares) for two different rewiring regimes. Insets: Linear plots of the distributions in their dominating degree range; comparison of steady-state prevalence \([I]_{MC}\) taken from MC simulations and \([I]_{NC}\) computed by the NC framework. Solid lines are predictions by the node cycle. Parameters \(p=0.008\), \(r=0.005\).
(a): \(w=0.025\), \(\tilde{w}=0.12\), \(\tilde{p}_S=0.044\), \(\tilde{p}_I=0.049\). (b): \(w=0.050\), \(\tilde{w}=0.22\), \(\tilde{p}_S=0.042\), \(\tilde{p}_I=0.045\). Cutoff for overall degree in NC matrices is \(k_\text{max}=80\).
MC simulations according to \cite{gillespie} with \(N=5\e4\) nodes, \(\langle k\rangle=7\), initial Erd\H{o}s-R\'{e}nyi graph and initial prevalence \([I]_0=0.6\). Statistics were recorded at \(t=3\e4\) for \(10^3\) network realizations. Error bars are smaller than markers.
}\label{f:DD}
\end{figure}

\begin{figure}
\centering
  \includegraphics[width=8.6cm]{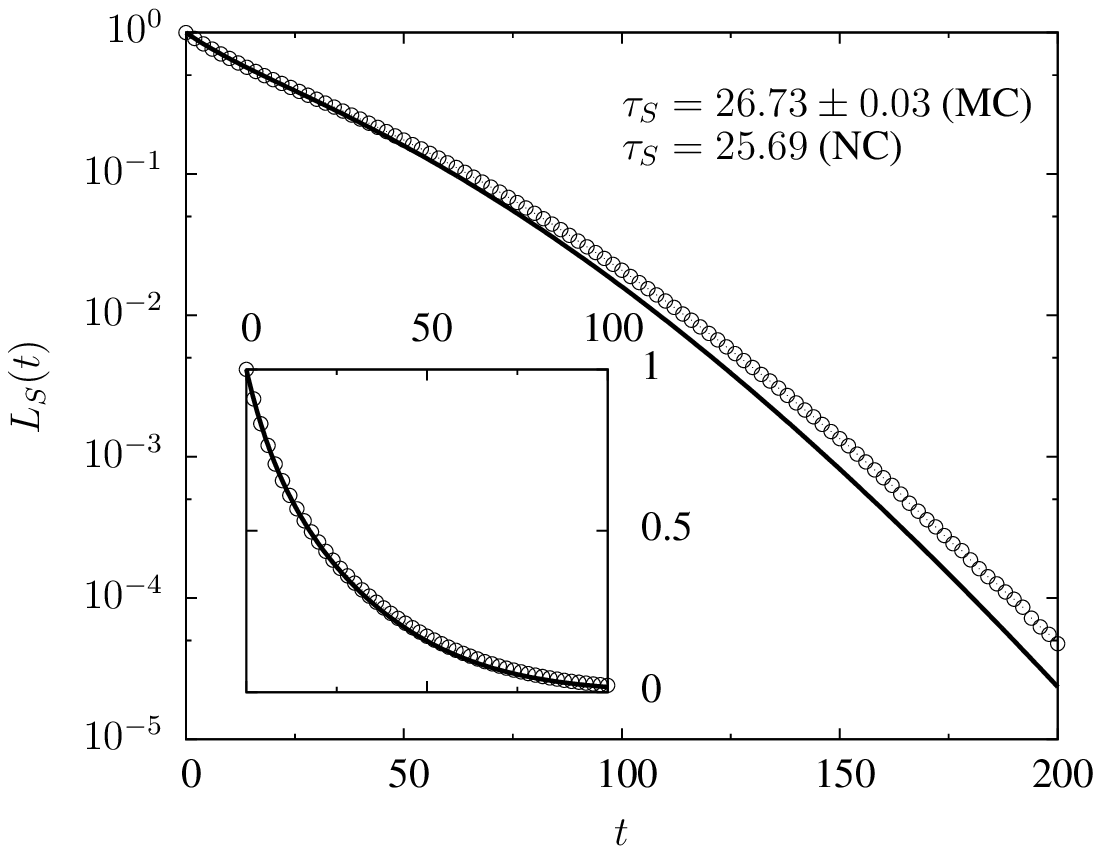}
  \caption{Survival function of S-nodes in steady state. Inset: Linear plot of the distribution for most prevalent lifetimes}; comparison of steady-state mean lifetimes \(\tau_S\) taken from MC simulations and computed by the NC framework. The solid line is the prediction by the node cycle. Parameters and initial conditions as in Fig. 2(b). Error bars are smaller than markers.
\label{f:LT}
\end{figure}

Computing the NC stationary state and choosing  values of the correspondence parameters such that the consistency conditions (i) and (ii) hold, the NC description can be mapped onto the stationary state of a particular network model.
However, the constraints set by (i) and (ii) may be impossible to satisfy exactly.
Because of the mean field approximation involved in the assumption of constant \(\tilde{p}_S\) and \(\tilde{p}_I\), the stochastic process we have dealt with does not capture the weak correlations present in the network. This translates into (i) and (ii)  not being exactly fulfilled, and so the description of the stationary state given by the NC is only approximate. 
In order to compare the analytic results with simulations on networks, we have used the free parameters \(\tilde{w}\), \(\tilde{p}_S\) and \(\tilde{p}_I\) to minimize the squared distance of the NC output to the target values set by conditions (i) and (ii), bringing the stochastic model to represent as accurately as possible the state and degree evolution of a network node. 
We shall refer to the analytic stationary degree distributions obtained through the method outlined above for the optimal choice of \(\tilde{w}\), \(\tilde{p}_S\) and \(\tilde{p}_I\) as the NC degree distributions. 
Results for a high and a low rewiring regime in a network of average degree \(\langle k\rangle = 7 \) are presented in Fig.~\ref{f:DD}. 
For different choices of \(w\), \(r\) and \(p\) in the stationary active phase,
the NC degree distributions as well as \(\Phi_I^*(x,y)\) show very good quantitative agreement with those obtained from
long Monte Carlo (MC) simulations of the network model of \cite{Gross} and \cite{quebec}, as do cumulative node lifetime distributions as given by Eq.~(\ref{e:LT}) (Fig.~\ref{f:LT}).
Our method predicts, through combinatorial handling of \(P^*_{S,I}(x,y)\), steady-state densities of any star motif. With steady-state node densities being given through
\begin{equation*}
\tau_S/\tau_I=(1-[I]_{NC})/[I]_{NC} , 
\end{equation*}
the density of triplets with a central I-node connected to a S- and another I-node, normalized by the system size, is illustratively computed as
\begin{equation*}
[IIS]_\mathrm{NC}=\sum\limits_{x,y=0}^{\infty}\frac{x\,y}{[I]_{NC}}P_S^*(x,y) \,.
\end{equation*}
For the parameters of Fig.~\ref{f:DD}(a), this yields \([IIS]_\mathrm{NC}=3.844\), with \([IIS]_\mathrm{MC}=3.824\pm0.004\) from MC simulations.
The small discrepancies (see Fig.~\ref{f:DD}) between steady-state moment densities
observed in MC simulations 
and those predicted by the NC method are another consequence of the constant \(\tilde{p}_S\) and constant \(\tilde{p}_I\) approximation. An improvement on this approximation would be to consider 
\(\tilde{p}_S\) and \(\tilde{p}_I\) to be linear functions of time chosen to represent status and degree correlations of the network model. For instance,  \(\tilde{p}_I\) should increase in time along the I-stage, because older (lower degree) I-nodes 
have susceptible neighbors that are younger than average S-nodes, 
 and younger S-nodes  
in turn have more than the average number of infected neighbors. Although it would no longer be possible in this case to find a simple solution like Eq.~(\ref{e:Fs}) for the generating functions \(F_{S,I}\), the whole NC procedure could in principle be carried out numerically for \emph{any} set of positive rate functions to yield an irreducible transition matrix \(\hat{P}\), a unique stationary degree distribution \(P^*_{S,I}(x,y)\) for each stage, and improved approximations for all the moment densities that characterize the network's steady state.

\section{Discussion and conclusions}

Apart from giving a complete description of the network's steady state that is in excellent quantitative agreement with the results of MC simulations, the NC method also shows that convergence of an arbitrary ICD to a unique steady-state topology determined by the parameters 
\(w\), \(r\), \(p\) and by the network's fixed degree will take place once the rate functions
associated with the correspondence parameters \(\tilde{w}\), \(\tilde{p}_S\) and \(\tilde{p}_I\)
are defined. These may be simply chosen as the constants that optimize the consistency conditions (i) and (ii), or, as mentioned above, may be taken as functions of time  with more free parameters, allowing a closer match of the NC model with the network-based system. In either case,
\(\hat{P}\) encodes a time-homogeneous Markov chain, meaning that the method applies only to the network model's stationary state and not to its transients. However, the correspondence parameters embody the mean-field approximations involved in the NC construction, and in the simplest case they are related only with two steady-state moment densities, \([SI]\) and \([I]\). If, as predicted in \cite{Gross} and confirmed by simulations, in the stationary active phase these densities reach equilibrium independently of network structure, then the NC approach shows that this implies that the network topology, as described by various probability distributions, will reach equilibrium as well.

Hence the pairwise model of \cite{Gross} and our NC approach should be considered complementary analytic frameworks to describe approximately adaptive networks. The former allows for a complete \emph{global} treatment of system dynamics, at the cost of confining the level of description to low-order moment \emph{densities}. The latter is a \emph{local} model, in that it identifies active steady states and gives a comprehensive description of their equilibrium dynamics through various characteristic \emph{distributions}.

In conclusion, taking the SIS network model with rewiring of Gross et al. as an example, we developed a self-sufficient method 
to derive analytically the steady-state 
subensemble degree distributions, and other measures that characterize the unique steady state in the simple endemic phase of the model. The only requirements of the construction are that we have 
a cyclic two-state
system, that the state changes are due to node 
or contact processes with constant rates, and that the network dynamics is due to link processes also with constant rates. The requirement of constant rates may 
be relaxed at the expense of an analytic solution for the probability generating functions not being available in this case.  As for the improved approximations briefly discussed in the previous section, 
dealing with non-linear transition rates would require the whole NC procedure to be carried out
numerically to yield the stationary degree distribution for each stage.
The requirement of a two-state system may also be relaxed to include cyclic multi-state systems such as SIRS or
rock-paper-scissors dynamics \cite{gross2011}. 
The method developed here can therefore serve
as a tool to explore the relation between the node's status change and rewiring rules
and the resulting network structure in different contexts, ranging from infection to opinion dynamics. 

\acknowledgments
The authors would like to thank Andrea Parisi for many helpful discussions.
Financial support from the Portuguese Foundation for Science and Technology (FCT) under Contract POCTI/ISFL/2/261 is gratefully acknowledged. The first and second authors were  also supported by FCT under Grant No. SFRH/BD/45179/2008 (S.W.) and CFTC-618-BII-02/08 (T.A.).


\begin{thebibliography}{0}

\bibitem{general_refs2}
\Name{P. Holme \and M. E. J. Newman} 
\REVIEW{Phys. Rev. E}{74}{2006}{056108};
\Name{N. H. Fefferman \and K. L. Ng} 
\REVIEW{Phys. Rev. E}{76}{2007}{031919};
\Name{S. Van Segbroeck, F. C. Santos, T. Lenaerts \and J. M. Pacheco} 
\REVIEW{Phys. Rev. Lett.}{102}{2009}{058105};
\Name{A. Szolnoki \and M. Perc} 
\REVIEW{EPL}{86}{2009}{30007}.

\bibitem{general_refs1}
\Name{M. G. Zimmermann, V. M. Egu\'\i luz \and M. San Miguel} 
\REVIEW{Phys. Rev. E}{69}{2004}{065102};
\Name{O. Gr\"{a}ser, C. Xu \and P. M. Hui} 
\REVIEW{EPL}{87}{2009}{38003};
\Name{F. Vazquez, V. M. Egu\'\i{}luz \and  M. San Miguel} 
\REVIEW{Phys. Rev. Lett.}{100}{2008}{108702};
\Name{G. I\~niguez, J. Kert\'esz, K. K. Kaski \and R. A. Barrio}  
\REVIEW{Phys. Rev. E}{80}{2009}{066119}.

\bibitem{Barabasi-Albert}
\Name{A.-L. Barab\' asi \and R. Albert} 
\REVIEW{Science}{286}{1999}{509}.

\bibitem{Murray}
\Name{J. D. Murray}\Book{Mathematical Biology}\Publ{Springer, New York}
\Year{1993}.

\bibitem{bondperc}
\Name{P. Grassberger}  
\REVIEW{Math. Biosci.}{63}{1983}{157};
\Name{M. E. J. Newman}  
{Phys. Rev. E}{66}{2002}{016128}; 
\Name{L. M. Sander, C. P. Warren \and I. M. Sokolov}  
\REVIEW{Physica A}{325}{2003}{1}.

\bibitem{Gross} 
\Name{T. Gross, C. J. Dommar D'Lima \and B. Blasius}
\REVIEW{Phys. Rev. Lett.}{96}{2006}{208701}.

\bibitem{Gross2}
\Name{T. Gross \and I. G. Kevrekidis} 
\REVIEW{EPL}{82}{2008}{38004}.

\bibitem{adaptiveSIX}
\Name{B. Bagnoli, P. Li\'o \and L. Sguanci} 
\REVIEW{Phys. Rev. E}{76}{2007}{061904};
\Name{L. B. Shaw \and I. B. Schwartz} 
\REVIEW{Phys. Rev. E}{77}{1008}{066101};
\Name{S. Risau-Gusman \and D. H. Zanette} 
\REVIEW{J. Theor. Biol.}{257}{2009}{52};
\Name{S. Funk, E. Gilad, C. Watkins \and V. A. A. Jansen} 
\REVIEW{Proc. Natl. Acad. Sci. USA}{106}{2009}{6872};
\Name{L. B. Shaw \and I. B. Schwartz} 
\REVIEW{Phys. Rev. E}{81}{2010}{046120};
\Name{C. Lagorio, M. Dickison, F. Vazquez, L. A. Braunstein, P. A. Macri, M. V. Migueles, S. Havlin and H. E. Stanley} 
\REVIEW{Phys. Rev. E}{83}{2011}{026102}.

\bibitem{quebec}
\Name{V. Marceau, P. A. No\"el, L. H\'ebert-Dufresne, A. Allard \and L.~J. Dub\'e} 
\REVIEW{Phys. Rev. E}{82}{2010}{036116}. 

\bibitem{vanKampen}
\Name{N. G. van Kampen}\Book{Stochastic Processes in Physics and Chemistry}\Publ{Elsevier, Amsterdam}
\Year{1981}.

\bibitem{gillespie}
\Name{D. T. Gillespie}
\REVIEW{J. Comput. Phys.}{22}{1976}{403}.

\bibitem{gross2011}
\Name{G. Demirel, R. Prizak, P. N. Reddy \and T. Gross} 
\Review{Eur. Phys. J. B}\Year{2011}
\doi{http://dx.doi.org/10.1140/epjb/e2011-10844-4}


\end{thebibliography}
\end{document}